\documentclass[10pt,twocolumn,twoside]{IEEEtran}

\IEEEoverridecommandlockouts    
\usepackage{psfrag}
\usepackage{amsmath,amssymb,amsfonts,bbm,nicefrac,mathtools,lipsum}
\usepackage{latexsym}
\usepackage{graphicx}
\usepackage{balance}
\usepackage{colortbl}
\usepackage{fancyhdr}
\usepackage{epstopdf}
\usepackage{float}
\usepackage{hyperref}
\usepackage{booktabs,tabularx}
\usepackage{algpseudocode} 

\usepackage{soul}
\usepackage[ruled]{algorithm}

\graphicspath{{./figures/}}

\date{ }
\newtheorem{theorem}{\textbf{Theorem}}

\begin{document}

\title{A Model Predictive Control Approach for Quadrotor Cruise Control}

\author{Zekai (David) Chen (5705762), Leon Kehler (4834186)
\thanks{Leon Kehler and Zekai Chen are master's students at TU Delft, DCSC, The Netherlands. 
E-mail addresses: \texttt{Z.Chen-58, L.Kehler}\ \texttt{@student.tudelft.nl}. \smallskip
}
}
\maketitle         

% How to find "Some Application":
% 1. http://ieeexplore.ieee.org
% 2. Search "model predictive control" from 201x on
% 3. Restrict search to "Journals & Magazines"
% 4. Restrict search to the "Publication Title" (titles) of interest (e.g. Intelligent Transportation Systems, Power Systems, Power Electronics, ...)

\begin{abstract}
This paper investigates the application of a Model Predictive Controller (MPC) for the cruise control system of a quadrotor, focusing on hovering point stabilization and reference tracking. Initially, a full-state-feedback MPC is designed for the ideal scenario. To account for real-world conditions, a constant disturbance is introduced to the quadrotor, simulating a gust of wind in a specific direction. In response, an output-feedback offset-free MPC is developed to stabilize the quadrotor while rejecting the disturbance. We validate the design of the controller by conducting stability analysis, as well as numerical simulations under different circumstances. It is shown that the designed controller can achieve all the expected goals for the cruise control, including reference tracking and disturbance rejection.\\

This project was implemented using Python and the CVXPY library for convex optimization~\cite{cvxpy}.
\end{abstract}

% Some Application is ... relevant for... The standard control strategy for Some Application is... which... cannot handle operational constraints... and/or may result in... non-optimal performance...

% The systems dynamics are linear/nonlinear, ... where ... are the system states, ... are the control inputs, and ... are the system outputs.

\section{Introduction} \label{sec:intro}
% 1/2-1 page

% How does a micro air vehicle find its way in a building?

% Identify the robotic system and the associated mathematical model for its kinematics, dynamics
% and motion equation. DONE 
% Identify the workspace and configuration space in which the robot operates. DONE
% Describe the chosen algorithm for motion planning and control of the mobile robot. You will
% also justify your choice.
% Design, implement and evaluate the chosen motion planner to solve a motion-planning task and
% navigate a mobile robot.
% Describe the latest developments in planning and decision-making for mobile robots. This
% includes the comparison between different methods

The main objective of this project is to develop a model predictive controller (MPC) that integrates state, input, and system dynamic constraints to achieve quadrotor stabilization. Consequently, a full-state-feedback MPC and an output-feedback offset-free MPC are designed and implemented in \autoref{sec:MPC}. The stability analysis presented in \autoref{sec:Stability} aims to validate the controller designs. Additionally, \autoref{sec:Simulation} explores the behavior of the MPC through several numerical simulations, including the examination of changing hyperparameters, comparison with the finite horizon unconstrained Linear Quadratic Regulator (LQR), disturbance rejection, and analysis of the different behaviors between nonlinear and linearized systems. The report concludes with a discussion and summary in \autoref{sec:Conclusion}. Lastly, the algorithm that is used to estimate the terminal set $\mathbb{X}_f$ and the process of obtaining the gain of finite horizon LQR is shown in the appendix.\\

This section introduces the nonlinear dynamics of the quadrotor. Further, it explains how the dynamics were linearized.

\begin{figure}[h!]
    \centering
    \includegraphics[width=0.75\columnwidth]{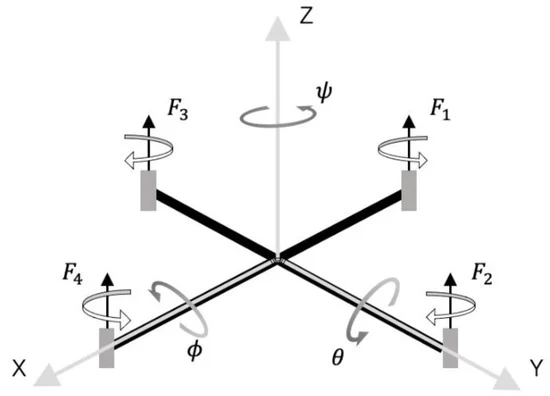}
    \caption{Kinetic diagram of the quadrotor~\cite{drone_diagram}}
    \label{fig:quadrotor}
\end{figure}

\subsection{Quadrotor Dynamics}
Figure \ref{fig:quadrotor} illustrates a kinetic diagram of the quadrotor. The dynamics of the quadrotor can be described by defining state variables that include cartesian coordinates $(X, Y, Z)$ and angular coordinates represented by roll, pitch, and yaw angles $(\phi, \theta, \psi)$, respectively. Furthermore, the inputs of the system is the total thrust of all motors and a torque around each axis. Consequently, this six-degree-of-freedom nonlinear system can be expressed in the following standard form:

\begin{align}\label{nonlinear_ss}
    \dot{x}&=f(x,u) \\
     y&=h(x,u).
\end{align}
\normalsize

The equations of motion of the nonlinear dynamic model of the quadrotor can be constructed using the Euler-Lagrange equations as shown in~\cite{ahmad2020simulation}.
\begin{align}
\label{eq:drone_cont_dynamics}
\begin{split}
\centering
    m\ddot{X}&= F(\cos{\phi}\sin{\theta}\cos{\psi}+\sin{\phi}\sin{\psi})\\
    m\ddot{Y}&= F(\cos{\phi}\sin{\theta}\sin{\psi}+\sin{\phi}\cos{\psi})\\
    m\ddot{Z}&= F\cos{\phi}\cos{\theta}-mg\\
    I_x\ddot{\phi}&=T_xl+\dot{\theta}\dot{\psi}(I_y-I_z)\\
    I_y\ddot{\theta}&=T_yl+\dot{\psi}\dot{\phi}(I_z-I_x)\\
    I_z\ddot{\psi}&=T_zl+\dot{\phi}\dot{\theta}(I_x-I_y)
\end{split}
\end{align}
\normalsize
with $F=F_1+F_2+F_3+F_4$. 

As a quick overview of the parameters, $F_1$, $F_2$, $F_3$ and $F_4$ are the upward thrust forces generated by the angular speeds $\omega_1$, $\omega_2$, $\omega_3$ and $\omega_4$ of the respective DC motors. $T_x$, $T_y$, and $T_z$ are the torques around their respective axis. \autoref{tab:parameter} lists the symbols, values, and meanings of the remaining parameters used in the model.
\begin{table}[h!]
\caption{Quadrotor Model Parameters}
    \label{tab:parameter}
    \begin{tabular}{|c|c|c|p{0.26\textwidth}|}
    \hline 
    Params & Value & Unit & Meaning \\ \hline
    $m$ & 1 & kg & Mass \\
    $g$ & 10 & m/s$^2$ & Gravitational acceleration \\
    $l$ & 0.2 & m & Arm length of the quadrotor frame \\ 
    $I_x$ & 0.11 & kg/m$^2$ & Moment of inertia about $x$ axis \\
    $I_y$ & 0.11 & kg/m$^2$ & Moment of inertia about $y$ axis \\
    $I_z$ & 0.04 & kg/m$^2$ & Moment of inertia about $z$ axis \\ \hline
    \end{tabular}
    
\end{table}

\subsection{Linearized Dynamics}
The design of the Model Predictive Controller (MPC) in this report relies on the Linear Time-Invariant (LTI) system. Hence, the above described system needs to linearized to fit the linear state-space description shwon in \autoref{eq:lin_ss}
\begin{align}
\label{eq:lin_ss}
\begin{split}
    \dot{x}&=Ax+Bu \\
    y &= Cx + Du
\end{split}
\end{align}  

To obtain the state-space model depicted above, the definition of the state variables $x$ is expanded to include not only positional and angular information but also their corresponding velocities. Furthermore, it is assumed that the system outputs its full state vector. Consequently, $x$ becomes a state vector composed of twelve entries, $u$ is an input vector composed of torques and forces, and $y$ is an output vector composed of a subset of the states. This definition of state variable vector $x$, input variables vector $u$, and output variables vector $y$ are shown below.
\begin{align*}
\centering
    x &=\left[
    X\,\ Y \,\ Z \,\ \phi \,\ \theta \,\ \psi \,\ \dot{X} \,\ \dot{Y} \,\ \dot{Z} \,\ \dot{\phi} \,\ \dot{\theta} \,\ \dot{\psi}
    \right]^T \\
    u &=\begin{bmatrix}
    F&T_x&T_y&T_z
    \end{bmatrix}^T \\
    y &= \begin{bmatrix}
    X&Y&Z&  \psi
    \end{bmatrix}^T 
\end{align*}
Next, linearizing around the hovering operating point, where the state vector is the zero vector, the torque inputs are zero, and only the thrust is equivalent to the quadrotor's weight ($m\cdot g$). The state-space matrices are hence obtained by differentiation as shown in \autoref{eq:linearization}. The differentiation was done symbolically in Python using CASADI.
\begin{align}\label{eq:linearization}
    A = \left.\begin{matrix}
\tfrac{\partial{f(x,u)}}{\partial{x}}
\end{matrix}\right|_{(0,mg)}, \quad B = \left.\begin{matrix}
\tfrac{\partial{f(x,u)}}{\partial{u}}
\end{matrix}\right|_{(0,mg)} \nonumber\\
    C = \left.\begin{matrix}
\tfrac{\partial{h(x,u)}}{\partial{x}}
\end{matrix}\right|_{(0,mg)}, \quad D = \left.\begin{matrix}
\tfrac{\partial{h(x,u)}}{\partial{u}}
\end{matrix}\right|_{(0,mg)} 
\end{align}
\normalsize

Finally, the system's dynamic model is discretized using a zero-order hold approach, with a sampling interval ($dt$) of 0.1 seconds. This discretization process is carried out utilizing the 'c2d' function from the Python control package, yielding the following dynamic model:
\begin{align}
x_{k+1} &= \Phi x_k + \Gamma u_k \\
y_k &= C x_k + D u_k
\end{align}
\normalsize
subsequently, the state variables vector for the discrete system becomes
\begin{align*}
\centering
    x &=\left[
    X\,\ Y \,\ Z \,\ \phi \,\ \theta \,\ \psi \,\ \Delta{X} \,\ \Delta{Y} \,\ \Delta{Z} \,\ \Delta{\phi} \,\ \Delta{\theta} \,\ \Delta{\psi}
    \right]^T
\end{align*}

% Parameter table
% \begin{table}[h!]
% \centering
% \begin{tabular}{|c|p{0.3\textwidth}|c|}
% \hline
% Symbol & Meaning & Value \\ \hline
% $\lambda$ & Healthy CD4+ helper-T cells generation rate & 1.0 \\ 
% $d$ & Natural death rate of uninfected CD4+ helper-T cell & 0.1 \\
% $\beta$ & Rate of transmission or infection & 1.0 \\
% $\eta$ & Effectiveness of antiretroviral therapy in reducing the infection rate & 0.9799 \\
% $a$ & Death rate of infected cells & 0.2 \\
% $b_1$ & The death rate of the first population of immune effector cells & 0.1 \\
% $b_2$ & The death rate of the second population of immune effector cells& 0.01 \\
% $c_1$ & Infected cells stimulate response rate for first immune effector cells & 0.03 \\
% $c_2$ & Infected cells stimulate response rate for second immune effector cells & 0.06 \\
% $p_1$ & First immune effector cells killing infected cells rate & 1.0 \\
% $p_2$ & Second immune effector cells killing infected cells rate & 1.0 \\
% $q$ & Viral dynamic & 0.5 \\
% $h$ & The death rate of the virus by the immune system & - \\ \hline
% \end{tabular}
% \caption{Parameter table}
% \label{ParametersTable}
% \end{table}

\section{Model Predictive Control Design} \label{sec:MPC}
Based on the dynamic model presented in Chapter \ref{sec:intro}, two MPCs are designed to stabilize the quadrotor at a certain hovering point. The two different designs were done aiming at two scenarios:
\begin{enumerate}
    \item A full information state feedback MPC for reference tracking for an undisturbed system.
    \item A output feedback offset-free MPC with Optimal Target Selection for reference tracking under a constant disturbance.
\end{enumerate}
The subsequent sections of this chapter will elaborate on the design details of these controllers, focusing specifically on cost functions, constraints, terminal sets, the observer, and other relevant aspects.

\subsection{Constraints}
% The state constraints, represented by the set $\mathbb{X}$, are defined such that limits are set for position and velocities, both linear and angular. The authors limit the linear velocities to an absolute maximum value of 10 m/s. As for the angular velocities, an absolute maximum of 0.5$\pi$ rad/s is set. The angles are also constrained as follows $\theta$,$\phi\in[-\pi/4,\pi/4]$ and $\psi\in[-\pi,\pi]$ due to physical considerations. Hence, these constraints can be represented in compact form as $|H x_k| \leq e$, where
% \begin{align*}
%     H &= 
%     \begin{bmatrix}
%         I_{3\times3}& 0_{3\times3} &0_{3\times3}&0_{3\times3}\\
%         0_{3\times3} &I_{3\times3} &0_{3\times3} & 0_{3\times3}\\
%          0_{3\times3} &0_{3\times3} & I_{3\times3}& 0_{3\times3}\\
%          0_{3\times3} &0_{3\times3} & 0_{3\times3}&I_{3\times3}\\
%     \end{bmatrix}\\
%     e&= [\infty; \infty; \infty;\pi/4;\pi/4;\pi;10; 10; 10; 0.5\pi; 0.5\pi; 0.5\pi]
% \end{align*} 

% The rotational speed of the motors at the hovering point can be obtained as follows
% \begin{align}
%     F&=mg\Leftrightarrow F=9.81\text{ N}\\
%     F&=K_t\textstyle\sum^4_{i=1}w_i^2\Leftrightarrow F=4K_tw_i^2\Leftrightarrow w_i = 904.16 \text{ rpm}
% \end{align}
% for $i=1,2,3,4$. Assuming the maximum rotational speed that each motor $i$ can achieve is $2\times w_i = 1808.32$ rpm, the maximum trust is set to $F_{max}= 39.24$ N.

Both MPC designs address the physical system's inherent constraints on actuators and sensors, which introduce limitations on the inputs and states during the optimization process. Specifically, the input constraints include the maximum total thrust the quadrotor can generate and the torque constraints. To be more specific, the rotational speed of the motors at the hovering point can be obtained as follows:
\begin{align}
    F_i&=mg\Leftrightarrow F_i=9.81\text{ N},\\
    F_i&=K_t\textstyle\sum^4_{i=1}w_i^2 \Leftrightarrow F_i=4K_tw_i^2\Leftrightarrow w_i = 904.16 \text{rpm}
\end{align}
\normalsize
for $i=1,2,3,4$. Assuming the maximum rotational speed that each motor $i$ can achieve is $2\times w_i = 1808.32$ rpm, the maximum trust is set to $F_{max}= 39.24$ N. Having these values, the control input constraints, represented by the set $\mathbb{U}$, can be defined as
\begin{align*}
-mg&=-9.81 \text{ N}\leq u_1 \leq F_{max}-mg=29.43 \text{ N},\\
   |u_2|& \leq lK_t(w_1^2-w_3^2)=1.47 \text{ Nm},\\
   |u_3| &\leq lK_t(w_2^2-w_4^2)=1.47 \text{ Nm},\\
   |u_4| &\leq K_d(w_2^2+w_4^2-w_1^2-w_3^2)=0.02 \text{ Nm}.
\end{align*}
\normalsize

Similar to the input constrict $\mathbb{U}$, the state constraints $\mathbb{X}$ follow from physical limits being imposed on the quadrotor model and its environment. It is assumed that the quadrotor operates in an open field, thereby imposing very broad constraints on its position $(X,Y,Z)$ with a range of 100 meters. However, to guarantee the validation of the drone dynamic, the roll and pitch angle of the rotor is constrained between $-\frac{\pi}{2}$ and $\frac{\pi}{2}$. The speed of the quadrotor is assumed to be bounded between $-2 m/s$ and $2 m/s$ for safety reasons. Additionally, all angular speeds are bounded between $-3 \pi/s$ to $3 \pi/s$ in accordance with the competition drone specifications. As a result, the input and state constraints can be represented in the form of

\begin{equation}
M = \begin{bmatrix}
-9.8 \\
-1.46 \\
-1.47 \\
-0.02
\end{bmatrix}
\leq
\begin{bmatrix}
F_k \\
T_{x,k} \\
T_{y,k} \\
T_{z,k}
\end{bmatrix}
\leq
\begin{bmatrix}
29.43 \\
1.47 \\
1.47 \\
0.02
\end{bmatrix}
= T
\end{equation}
\normalsize and 
\[
-100 \leq X,Y,Z \leq 100
\]
\[
-3 \leq \Delta_X,\Delta_Y,\Delta_Z \leq 3
\]
\[
\frac{-\pi}{2} \leq \phi, \theta \leq \frac{\pi}{2}
\]
\[
-2\pi \leq \psi \leq 2\pi
\]
\[
-3\pi \leq \Delta_\phi, \Delta_\theta, \Delta_\psi \leq 3\pi
\]

These above constraints can be compactly combined as $G u(k) \leq g$ and $E x(k) \leq e$ with definition of $G$ and $g$:

\[
G = \begin{bmatrix}
I \\
-I
\end{bmatrix}, \quad
g = \begin{bmatrix}
T \\
-M
\end{bmatrix}
\]
\normalsize 
The same method of constructing constraints can be applied to state constraints when applying $E$ and $e$. Such form makes the feasible set polyhedral, which is important when estimating terminal set $\mathbb{X}_f$. Moreover, this makes the optimization problem a quadratic programming problem with linear constraints, which is computationally faster than quadratic constraints. 
% Subsequently, by augmenting the original state $x$ to $x_{aug} = [x,x,u,u]^T$, the state and input constraints can be compactly written as \autoref{eq:CompactConstraints} demonstrated, a form that is especially useful for estimating the terminal constraint set $\mathbb{X}_f = \left\{ x \in \mathbb{R}^n \mid Hx \leq h \right\}$.

% \begin{equation} \label{eq:CompactConstraints}
%     \begin{bmatrix}
%         E \\G 
%     \end{bmatrix}
%     x_{aug} \leq
%     \begin{bmatrix}
%         e \\
%         g
%     \end{bmatrix}
% \end{equation}

\subsection{Cost Functions}
The general definition of the cost function for an MPC with a prediction horizon $N$ can be written as \autoref{eq:MPC_cost}.
\begin{equation} \label{eq:MPC_cost}
    V_N(x_0, \mathbf{u_N}) = \sum^{N-1}_{k=0}l(x_k,u_k)+V_f(x_N)
\end{equation}
The cost function shown in \autoref{eq:MPC_cost} composed of stage cost $l(x_k, u_k)$ and terminal cost $V_f(x_N)$. Both of these costs are defined in quadratic form, as illustrated in \autoref{eq:cost2}.
\begin{gather}  \label{eq:cost2}
    l(x_k,u_k)=\frac{1}{2}(x_k^TQx_k+u^T_kRu_k) \\
    V_f(x_N)=\frac{1}{2}x_N^TPx_N
\end{gather}
\normalsize
with matrix $Q$ and $R$ positive define diagonal matrixes. These two matrixes are also known as the weight matrixes. The matrix $P$ used for the definition of the terminal cost is the unique positive semi-definite solution to the Discrete Algebraic Riccati Equation (DARE). 

% By this design, the terminal cost $V_f(x)$ shares the same form as the unconstrained optimal cost $V^{uc}_{\infty}(x)$. This ensures that $x(N)$ will be in a neighborhood of the origin when $N$ is large enough. Subsequently, the optimal cost can take over and stabilize the system while respecting the state constraints.

\subsection{Terminal Set \(\mathbb{X}_f\)}
The terminal set $\mathbb{X}_f$ is designed to be the invariant constraint admissible set. Algorithm 1 adapted from~\cite{gilbert1991linear} is used to determine the maximal constraint admissible set for the regulation problem. In the work of Gilbert and Tan, an algorithm is proposed to recursively estimate the output admissible set $O_{\infty}$ by using the finitely determined property~\cite{gilbert1991linear}. Moreover, it is also found that when constraints set that consists of input and state constraints, but no cross terms, $\mathbb{Z}=\mathbb{U} \times \mathbb{X}$ is a polyhedron, the determination of $O_{\infty}$ is straightforward by solving a sequence of linear programming problems~\cite{gilbert1991linear}. Following a similar logic, this algorithm is adjusted in this work by replacing the $C(\Phi-\Gamma K)^t$ with the closed-loop system matrix $[K; I](\Phi-\Gamma K)^t$. This replacement changes the mapping from $x \rightarrow y$ to $x_{aug} \rightarrow x_{aug}$ with $x_{aug}$ containing not only the state variables but also the inputs, resulting in obtaining a state and input admissible invariant set instead of an output admissible set.

The algorithm presented in the pseudocode \ref{alg:estimate_set_Xf} demonstrates that the algorithm iteratively expands the set $\mathbb{X}_f$ by increasing $t$ until the set of all initial states $x$ for which the system can be controlled within the constraints. The result is another polyhedron $\mathbb{X}_f = \left\{ x \in \mathbb{R}^n \mid Hx \leq h \right\}$ that contains 480 constraints.

\subsection{Full Information State-Feedback MPC}
The full information state-feedback MPC makes use of the discrete system introduced in \autoref{sec:intro} and the availability of all the states, forming an optimal control problem $\mathbb{P}_N(x_0, t)$ shown in \autoref{eq:OptimalControlProb}.

\begin{equation} \label{eq:OptimalControlProb}
\begin{aligned}
\min_{u_N} \quad & V_N(x_0, u_N) \\
\text{s.t. } & x_{k+1} = \Phi x_k + \Gamma u_k \\
             & y_k = C x_k + D u_k \\
             & x_k \in \mathbb{X}, u_k \in \mathbb{U}\\
             & x_N \in \mathbb{X}_f
\end{aligned}
\end{equation}

The prediction horizon is chosen as $N=10$. For the sake of conciseness, the full information reference tracking controller is omitted, as it is essentially a simplified version of the output controller with disturbance rejection. Therefore, the following section will focus on the design of the output controller with disturbance rejection. The matrix $Q$ is chosen as a diagonal matrix with the elements $(10,10,100,10,10,10,1,1,1,1,1)$ on the diagonal, primarily penalizing the altitude of the quadrotor since the main task is hovering. Similarly, the matrix $R$ is chosen as a diagonal matrix with the elements $(0.1,1,1,1)$ on the diagonal, ensuring that sufficient control input $u$ can be provided to the drone.

\subsection{Output-Feedback MPC and Disturbance Rejection}
The previous full information state-feedback is built based on the assumption that all states are available for control and no existence of disturbance. Namely the output matrix $C=I$, the output reference $y_{ref}=0$, and disturbance $d=0$. However, this is not always the case in practice. Due to the limitations of the Inertial Measurement Unit (IMU), direct measurements of some state variables may be impossible. Additionally, the external environment can introduce disturbances to the system. These challenges necessitate the construction of an output-feedback controller with state estimation for reference tracking and disturbance rejection.

We created a scenario where the quadrotor is equipped with sensors that can only measure the quadrotor's pose, i.e. position and attitude, but no velocities. Additionally, a constant wind gust in the $X$-direction is imposed on the quadrotor as a disturbance. As a result, the state space representation of the discrete system can be written as \autoref{eq:disturbanceSS} shows:
\begin{equation} \label{eq:disturbanceSS}
    \begin{aligned}
        x_{k+1} &= \Phi x_k + \Gamma u_k + \Gamma_d d_k + \omega_k \\
        y_k &= C x_k + C_d d_k + \nu_k
    \end{aligned}
\end{equation}
\normalsize
with $d$ denotes a constant and unknown disturbance exerted on $X$. $\omega$ and $\nu$ are two white noises representing state disturbance and measurement noise with $\mathbb{E}(\omega_k)=0$ and $\mathbb{E}(\nu_k)=0$. To add disturbance rejection properties to the closed-loop system and the ability to estimate the unknown disturbance, the original discrete system is augmented by taking disturbance $d$ as an auxiliary state, yielding the following state space representation in \autoref{eq:AugmentedSS}.

\begin{equation} \label{eq:AugmentedSS}
\begin{aligned}
\begin{bmatrix}
x_{k+1} \\
d_{k+1}
\end{bmatrix}
&=
\begin{bmatrix}
\Phi & \Gamma_d \\
0 & I
\end{bmatrix}
\begin{bmatrix}
x_k \\
d_k
\end{bmatrix}
+
\begin{bmatrix}
\Gamma \\
0
\end{bmatrix}
u_k
+
\begin{bmatrix}
w \\
w_d
\end{bmatrix} \\
y_k &=
\begin{bmatrix}
C & C_d
\end{bmatrix}
\begin{bmatrix}
x_k \\
d_k
\end{bmatrix}
+ v
\end{aligned}
\end{equation}
\normalsize
the matrix $\Gamma_d$ and $C_d$ are chosen in the form so that the constant disturbance $d$ is only added to the state of $X$, stimulating a wind guest at $X$ position. The original system $(\Phi, \Gamma)$ is tested to be controllable. To guarantee that the augmented system $(\tilde{\Phi}, \Tilde{C})$ is observable, the observability of the original system $(\Phi, C)$ is checked by \autoref{eq:checkObs}.
\begin{equation} \label{eq:checkObs}
\text{rank} \begin{bmatrix}
I - \Phi & -\Gamma_d \\
C & C_d
\end{bmatrix} = n + n_d = 12 + 1 = 13
\end{equation}

A Luenberger observer is designed to reconstruct the state based on the output information. The principle of designing an observer is to ensure the observer error converges faster to zero than the closed-loop system. For a discrete-time system, this is achieved by ensuring the poles of $\Phi - L C$ are closer to the original compared to those of the closed-loop system $\Phi - \Gamma K$. In this project, a Kalman filter is employed to derive the observer gain $L$, which guarantees adherence to the observer design principle. The process and measurement covariance matrices are set to $Q_{K} = I$ and $R_{K} = I$ respectively. Consequently, the mathematical formula of the Luenberger observer is shown in \autoref{eq:ObsCLSys}. 

\begin{equation} \label{eq:ObsCLSys}
\begin{aligned}
\begin{bmatrix}
\hat{x}_{k+1} \\
\hat{d}_{k+1}
\end{bmatrix}
&= 
\underbrace{
\begin{bmatrix}
\Phi & \Gamma_d \\
0 & I
\end{bmatrix}
}_{\tilde{\Phi}}
\begin{bmatrix}
\hat{x}_k \\
\hat{d}_k
\end{bmatrix}
+ 
\begin{bmatrix}
\Gamma \\
0
\end{bmatrix}
u_k
+ 
L
(y_k - 
\underbrace{
\begin{bmatrix}
C & C_d
\end{bmatrix}
}_{\tilde{C}}
\begin{bmatrix}
\hat{x} \\
\hat{d}
\end{bmatrix}
)
\end{aligned}
\end{equation}

Lastly, given that not all the states are directly measurable, only the output reference $y_{ref}$ can be given to the system for performing reference tracking. However, the MPC objective function depends on $x_{ref}$. Consequently, the Optimal Target Selection (OTS) problem, as presented in \autoref{eq:OTS}, must be solved to determine the reference $x_{ref}$ and $u_{ref}$ based on the given $y_{ref}$ and the best current estimation of the disturbance $\hat{d}$. Note that the disturbance estimate evolves according to the observer dynamics. Hence, the OTS must be solved online.
\begin{equation} \label{eq:OTS}
(x_{ref}, u_{ref})(\hat{d}, y_{ref}) \in\left\{\begin{array}{l}
\arg \min_{x_r, u_r} J(x_r, u_r) \\
\text {s.t.}\left[\begin{array}{cc}
I-\Phi & -\Gamma \\
C & 0
\end{array}\right]\left[\begin{array}{l}
x_r \\
u_r
\end{array}\right] \\ =\left[\begin{array}{c}
\Gamma_d \hat{d} \\
y_{ref}-C_d\hat{d}
\end{array}\right] \\
\left(x_r, u_r\right) \in \mathbb{Z} \\
C x_r+\hat{d} \in \mathbb{Y}
\end{array}\right.
\end{equation}
\normalsize
Based on the $x_{ref}$ and $u_{ref}$ acquired by the above OTS problem, the stage and terminal costs are adjusted accordingly as demonstrated in \autoref{eq:cost_sfb_MPC} to perform the reference tracking.

\begin{equation} \label{eq:cost_sfb_MPC}
    \begin{aligned}
        l(x_k,u_k)=\frac{1}{2}(x_k-x_{ref})^TQ(x_k-x_{ref})\\
        + \frac{1}{2}(u_k-u_{ref})^TR(u_k-u_{ref}) \\
    V_f(x_N)=\frac{1}{2}(x_N-x_{ref})^TP(x_N-x_{ref})
    \end{aligned}
\end{equation}  
\normalsize
The matrix $Q$, $R$, $P$, and the prediction horizon $N$ are set the same as the full information state-feedback MPC. Consequently, the output-feedback offset-free MPC is built by solving the optimal control problem presented in \autoref{eq:OptimalControlProb}. 

In summary, a quadratic cost function is utilized, and the terminal cost $V_f(x)$ is designed in the same form as $V^{uc}_{\infty}(x)$, i.e. $\frac{1}{2}x^T P x$ with $P$ being the solution of the DARE. The state constraints $\mathbb{X}$ and input constraints $\mathbb{U}$ are designed as a polyhedral, containing the origin. Moreover, the terminal set $\mathbb{X}_f$ is chosen as the invariant constraint admissible set, also containing the origin. This design makes $\mathbb{X}_f $ in the shape of a polyhedral, rather than ellipsoidal. The latter would be the result of designing $\mathbb{X}_f$ as the sublevel set of $V_f(x) = \frac{1}{2}x^T P x$ with $P$ not necessarily the solution of the DARE that satisfies the condition that set $\mathbb{X}_f := \{ x \in \mathbb{R}^n \mid V_f(x) \leq c \}$ with $c > 0$ ensures $\mathbb{X}_f \subseteq \mathbb{X}$ and $K \mathbb{X}_f \subseteq \mathbb{U}$. This design choice adopted by this report makes sure that the optimal finite-horizon value function $V^0_N$ is the same as the optimal infinite-horizon value function $V^{uc}_{\infty}$ within $\mathbb{X}_f$, resulting in the MPC behaving as the LQR. Subsequently, the polyhedral shape of $\mathbb{X}_f$ facilitates the formulation of the MPC problem as a QP (Quadratic Programming) problem with linear constraints, which is computationally less demanding than solving a QCQP (Quadratically Constrained Quadratic Programming) problem that the ellipsoidal terminal set $\mathbb{X}_f$ leads to~\cite{rawlings:mayne}.
%\section{Model predictive control design}

% We consider a receding horizon MPC strategy with control horizon $N=...$

% The state constraints are... hence can be represented in compact form as $F x(k) \leq e$, where $F$ is the matrix 
% $$ F = \left[
% \begin{matrix}
% * & * & 0 \\
% * & * & * \\
% * & * & *
% \end{matrix}
% \right]
%  $$
% and $e$ is the vector...

% Control input constraints...

% Output constraints...

% We design the stage cost function $\ell(x,u) = ...$, the terminal cost function $V_{\textup{f}}(x) = ...$, and the terminal set $\mathbb{X}_{\textup{f}} = ...$

\section{Asymptotic stability} \label{sec:Stability}
In this section, the asymptotic stability of the linearized closed-loop system for the previously designed MPC is proved.

The open loop system is controllable because its Kalman matrix $[\Phi, \Phi \Gamma, \dots, \Phi^{11} \Gamma]$ is full rank. The Lyapunov Stability Theory states that:
\begin{theorem}[Lyapunov Stability]
\label{Lyapunov}
Suppose V(.) is a Lyapunov candidate function for the system $x^+ = f(x)$, If $V(x)>V(x_e)$ and $\nabla V(x)f(x) <0$, $\forall x \in \mathbb{X} \setminus x_e$ the equilibrium $x_e$ is stable; here $x_e = 0$.
\end{theorem}

For an optimal control problem with finite horizon $N$, by choosing appropriate $l$, $V_f$, and $\mathbb{X}_f$, there exists a Lyapunov function $V(.)$ that subsequently proves the asymptotic stability of the origin. However, for the to-be-controlled system, $x^+=\Phi x + \Gamma u$ where inputs and states are subjected to $u \in \mathbb{U}$ and $x \in \mathbb{X}$, determining a global Lyapunov function (LF) is difficult if not impossible because of the constraints. Thus, a local LF together with an invariant set $\mathbb{X}_f$ is determined to prove the asymptotic stability of the origin.

To show the asymptotic stability, we must show that the optimal cost function $V_N^0$ is a local LF. Practically, to prove the origin is exponentially stable in the region of attraction $\mathcal{X}_N$ through Theorem 2.19 (asymptotic stability of the origin) and Theorem 2.21 (Lyapunov function and exponential stability) in the book~\cite{rawlings:mayne}, it remains to be shown that Assumption 2.2, 2.3, and 2.14 in the same book are satisfied, and $\mathbb{X}_f$ contains the origin in its interior~\cite{rawlings:mayne}. The assumptions are shown to be satisfied each by each as follows.

\begin{itemize}
    \item \textbf{Assumption 2.2} (Continuity of system and cost). The function $f:\mathbb{Z} \Rightarrow \mathbb{X}, l: \mathbb{Z} \rightarrow \mathbb{R}_{\geq0}$ and $V_f:\mathbb{X}_f \rightarrow \mathbb{Z}_{\geq0}$ are continuous, $f(0,0)=0$, $l(0,0)=0$ and $V_f(0)=0$.
\end{itemize}
This assumption aims to show the continuity of the system and the costs. From the \autoref{sec:intro}, it is shown that the linearized system dynamic function $f$ is continuous and satisfies $f(0,0)=0$. Furthermore, the quadratic form of stage cost $l(x,u)$ and terminal cost $V_f(x)$ also meet the requirement for non-negative cost value and $l(0,0)=0$, $V_f(0)=0$. Thus, \textbf{Assumption 2.2 is fulfilled.}

\begin{itemize}
    \item \textbf{Assumption 2.3} (Properties of constraint sets). The set $\mathbb{Z}$ is closed and the set $\mathbb{X}_f \subseteq \mathbb{X}$ is compact. Each set contains the origin. If $\mathbb{U}$ is bounded (hence compact), the set $\mathbb{U}(x)$ is compact for all $x \in \mathbb{X}$. If $\mathbb{U}$ is unbounded, the function $u \rightarrow V_N(x,u)$ is coercive, i.e., $V_N(x,u) \rightarrow \infty$ as $|u| \rightarrow \infty$ for all $x \in \mathbb{X}$
\end{itemize}
This assumption aims to show that the constrained sets are compact and bounded and contain the origin. The input constraint set $\mathbb{U}$ is closed and bounded as shown in \autoref{sec:MPC}, and therefore compact. Since state constraint set $\mathbb{X}$ is closed, $\mathbb{Z} = \mathbb{X} \times \mathbb{U}$ is also a closed set. Additionally, the origin is contained in all three sets according to the inequalities. 

The terminal constraint set is represented by $\mathbb{X}_f = \left\{ x \in \mathbb{R}^n \mid Hx \leq h \right\}$. It can be proven straightforwardly that the origin is in the set by showing the fulfillment of the inequality. Furthermore, $\mathbb{X}_f$ is estimated by gradually expanding the set with the increase of $t$ while ensuring the fulfillment of the state constraints, this guarantees that $\mathbb{X}_f \subseteq \mathbb{X}$. Hence, the property of $\mathbb{X}$ is inherited by $\mathbb{X}_f$, guaranteeing the compactness. Hence, \textbf{Assumption 2.3 is fulfilled.}

\begin{itemize}
    \item \textbf{Assumption 2.14} (Basic stability assumption). $V_f(.)$, $X_f$ and $l(.)$ have the following properties 
    \begin{enumerate}
    \item $\forall x \in \mathbb{X}_f$, $\exists u \in \mathbb{U}$, such that
    \begin{enumerate}
        \item $f(x,u) \in \mathbb{X}_f$
        \item $V_f(f(x,u)) - V_f(x) \leq - l(x,u)$
    \end{enumerate}
    \item $\exists \kappa_{\infty} \text{ functions }\alpha_1, \alpha_2$, such that
    \begin{enumerate}
        \item $l(x,u) \geq \alpha_1(|x|)$, $\forall x \in \mathcal{X}_N$, $\forall u$ s.t. $(x,u) \in \mathbb{Z}$
        \item $V_f(x) \leq \alpha_f(|x|)$, $\forall x \in \mathbb{X}_f$
    \end{enumerate}
\end{enumerate}
\end{itemize}
The Assumption of 2.14 can be divided into two parts, the goal of Part 1) is to show that the terminal set $\mathbb{X}_f$ is a control invariant set for system $f(x,u)$ and the Lyapunov decrease of $V_f(x)$, the latter is a sufficient condition for the optimal cost function $V_N^0$ to be a local CLF. The goal of Part 2) is to show that the stage cost and the terminal cost have a lower bound and upper bound respectively. 

For simplicity reasons, we first show the lower bound for $l(x,u)$, and the upper bound for $V_f(x)$ using knowledge from the linear algebra. For the quadratic form of stage cost $l(x,u)$, because matrix $Q$ and $R$ are semi-positive definite, the below relation can be obtained in \autoref{eq:l_lb}. 
\begin{equation} \label{eq:l_lb}
    l(x, u) \geq \frac{1}{2} x^TQx \geq \frac{1}{2} \lambda_{\min}(Q) |x|^2 = \alpha_1(|x|) \quad \forall x \in \mathcal{X}_N
\end{equation}
The same method can be used to obtain the upper bound of the terminal cost $V_f(x)$ shown in \autoref{eq:Vf_ub}. 
\begin{equation} \label{eq:Vf_ub}
    V_f(x) \leq \frac{1}{2} \lambda_{\max}(P) |x|^2 = \alpha_f(|x|) \quad \forall x \in \mathbb{X}_f
\end{equation}
Hence, \textbf{the Part 2) of Assumption 2.14 is fulfilled}.

\begin{figure}[h!]
    \centering
    \includegraphics[width=\columnwidth]{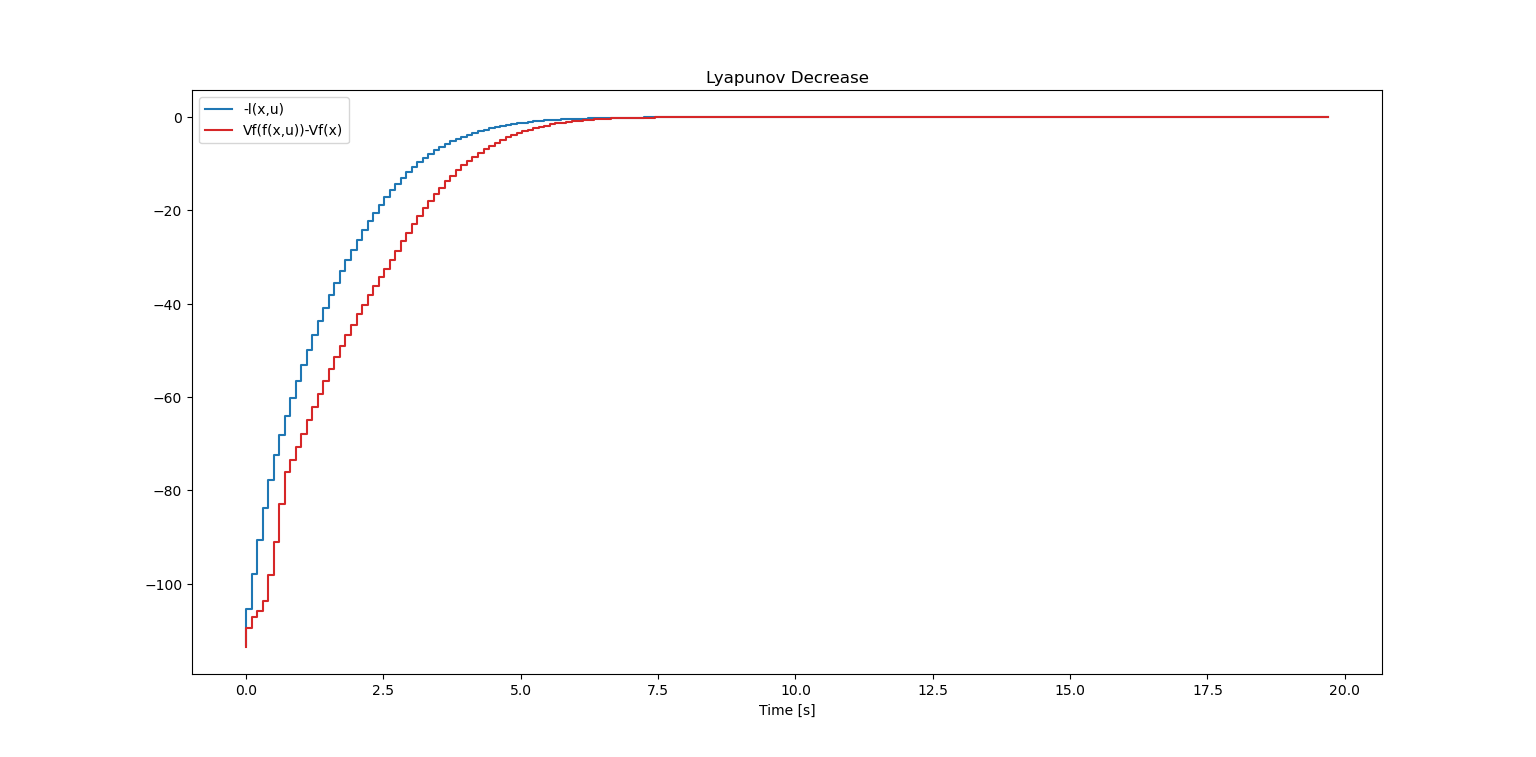}
    \caption{Lyapunov decrease with costs $l(x,u) = \frac{1}{2}(x^TQx+u^TRu)$ and $V_f(x) = \frac{1}{2}x^TPx$}
    \label{fig:LyapunovDecrease}
\end{figure}

In \autoref{sec:MPC}, $\mathbb{X}_f$ is estimated as the maximal constraint invariant admissible set, i.e., ensuring that for all $\forall x \in \mathbb{X}$, there exists an input $\exists u \in \mathbb{U}$ such that it keeps the system within the set while satisfying all constraints. This is confirmed by estimating using Algorithm \ref{alg:estimate_set_Xf}~\cite{gilbert1991linear}.
\begin{equation} \label{eq:LyapunovDecrease}
    V_f(f(x,u)) - V_f(x) \leq -l(x,u)
\end{equation}
Lastly, we show that $\forall x \in \mathbb{X}_f$, there exists $\exists u \in \mathbb{U}$ such that \autoref{eq:LyapunovDecrease} holds in \autoref{fig:LyapunovDecrease}. Hence, \textbf{Part 1) of Assumption 2.14 is fulfilled} and \textbf{Assumption 2.14 is fulfilled}.

Based on the above analysis, Assumptions 2.2, 2.3, and 2.14 are all fulfilled. Therefore, according to Theorem 2.19 and 2.21, the origin is exponentially stable in the region of attraction $\mathcal{X}_N$~\cite{rawlings:mayne}.
%\section{Asymptotic stability}

% In this section, we show that the designed MPC asymptotically stabilized the closed-loop system. With this aim, we verify the assumptions of Theorem... in the book~\cite{rawlings:mayne}

% Assumption 2.x: ... 

% Assumption 2.y: ...

\section{Numerical Simulations} \label{sec:Simulation}
In this chapter, a series of numerical simulations are conducted to demonstrate the efficacy of the designed Model Predictive Control (MPC). The performance of MPCs with varying hyper-parameters is compared for reference tracking. Additionally, the output feedback MPC is evaluated for its disturbance rejection capability. Furthermore, the superior performance of MPC compared to the unconstrained finite horizon Linear Quadratic Regulator (LQR) is illustrated. Finally, the model mismatch between the nonlinear and linearized system is investigated.

\subsection{Comparison of different $N$}
\begin{figure}[h!]
    \centering
    \includegraphics[width=\linewidth]{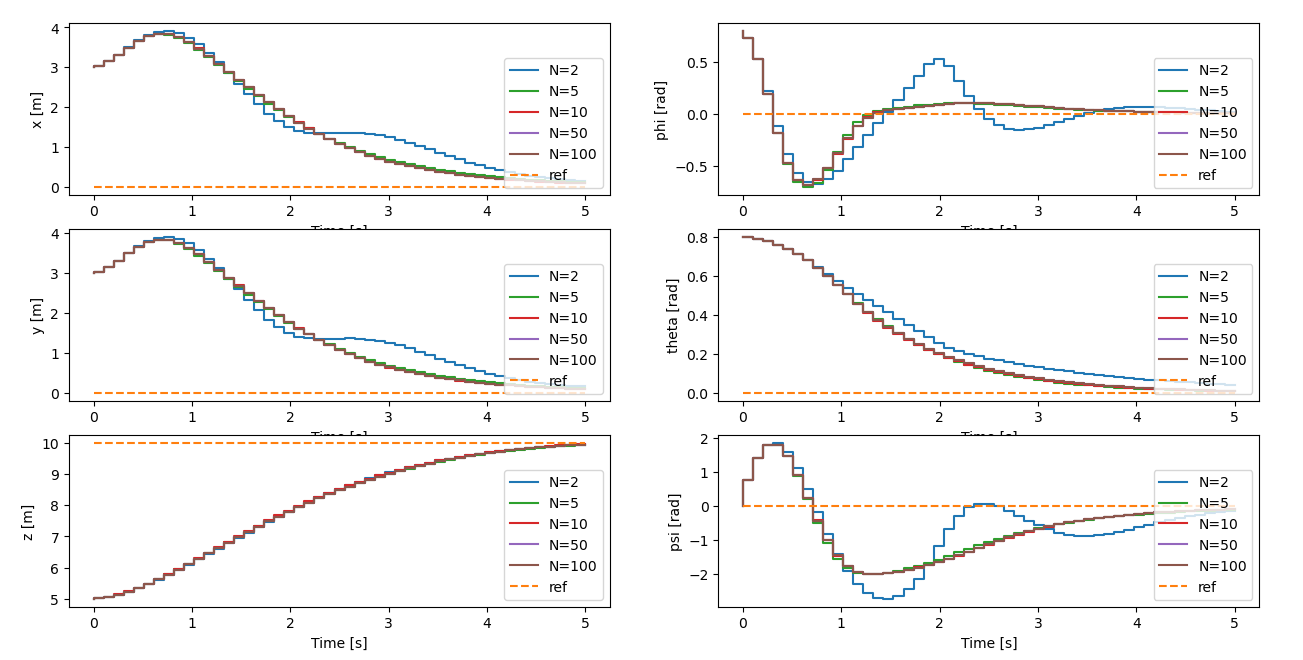}
    \caption{Result of MPC with different $N$}
    \label{fig:diffN}
\end{figure}

We first investigate the relationship between the performance of the MPC and the perdition horizon $N$. In the simulation, 5 different $N$ are tested ranging from 2 to 100. \autoref{fig:diffN} illustrated the corresponding simulation result of the first six states. To guarantee the fairness of comparison, only the prediction horizon $N$ is varied while other hyper-parameters are kept the same. \autoref{fig:diffN} reveals that the full-state feedback MPC is able to stabilize the quadrotor at the hovering point. However, more oscillations are noticed when $N$ is small. This can be explained by the limited information the controller can get when forming an optimal control policy. Moreover, when $N$ is large, despite the fact that the controller is able to stabilize the system, the computational time increases dramatically.

\begin{figure}[h!]
    \centering
    \includegraphics[width=\linewidth]{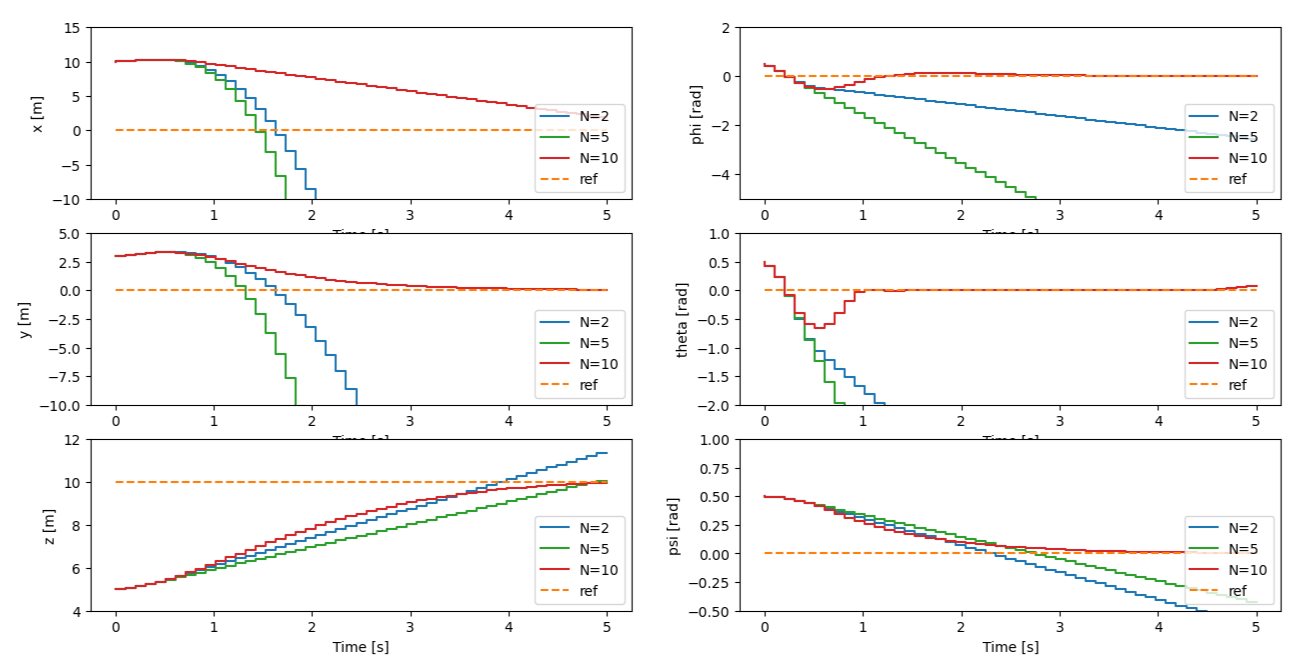}
    \caption{MPC with different $N$ and $x_0$ far away from $\mathbb{X}_f$}
    \label{fig:diffN_outX}
\end{figure}

The scenario in \autoref{fig:diffN} shows that the MPC is able to stabilize the system for different choices of $N$. This stabilization is achieved because the MPC successfully guides the state $x(N)$ into the control-invariant admissible set $\mathbb{X}_f$ within the horizon $N$. However, this outcome is not universally guaranteed. When the initial condition $x_0$ is too far away from $\mathbb{X}_f$, the MPC fails to steer the state $x$ into $\mathbb{X}_f$ within $N$ steps due to the small $N$, resulting in instability. This issue is illustrated in \autoref{fig:diffN_outX} where only the MPC with $N\geq10$ can stabilize the system. Hence, the above analysis indicates a need to find a trade-off between quality and computation in real-life applications.

The solving time $t_s$ of each iteration is affected by the choice of N. \autoref{tab:solving_time_N} shows that this is almost a linear relationship; the offset may be due to some initialization of CVXPY.

\begin{table}[h!]
\centering
\begin{tabular}{|c|ccccc|}
\hline
\textbf{N} & 2 & 5 & 10 & 50 & 100 \\ \hline
\textbf{t}$_s$ \textbf{[s]} & 0.02 & 0.03 & 0.06 & 0.26 & 0.51 \\ \hline
\end{tabular}
\caption{Solving time per iterate for different $N$}
\label{tab:solving_time_N}
\end{table}

\subsection{Comparison of different weight matrices $Q$ and $R$}
\begin{figure}[h!]
    \centering
    \includegraphics[width=\columnwidth]{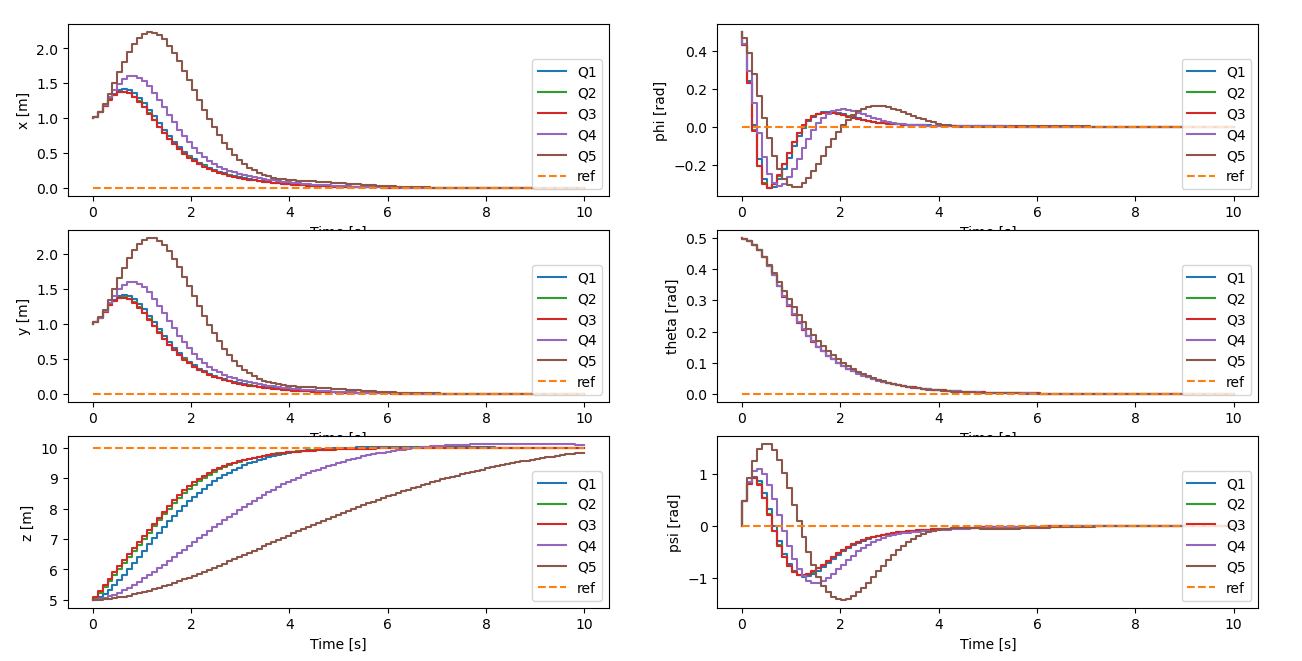}
    \caption{Result of MPC with different $Q$}
    \label{fig:diffQ}
\end{figure}

The relationship between the weight matrix in the cost function $Q$ and $R$, and the performance of the MPC is investigated further. The two matrices penalize state $x$ and input $u$ when solving the optimization problem. Based on the quadratic form of the cost function shown in \autoref{eq:MPC_cost}, it is expected that a larger value in the cost function $Q$ will result in a better performance for the MPC. \autoref{fig:diffQ} demonstrates the simulation results of different $Q$ while other hyper-parameters are kept identical. Among all the options, the controller performs the best when $Q$ is the largest, meaning states are penalized hard. The opposite performance can be seen in $Q_5$ when all the elements are deliberately chosen small.

\begin{figure}[h!]
    \centering
    \includegraphics[width=\columnwidth]{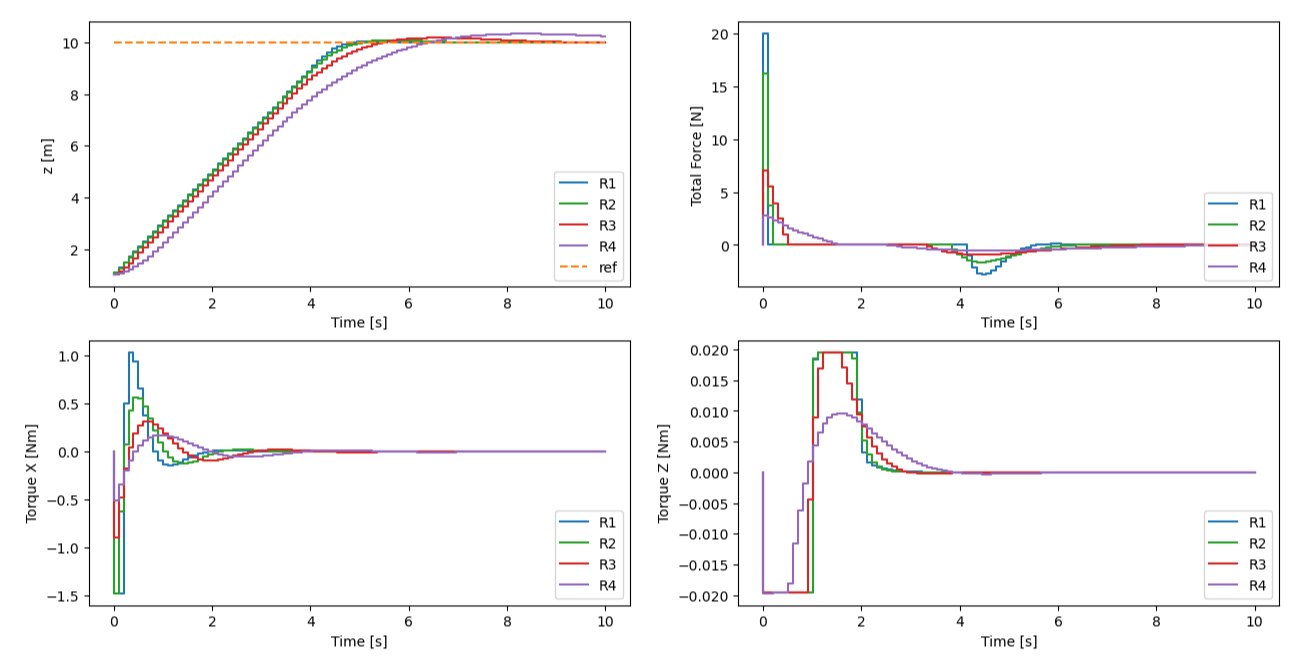}
    \caption{Result of MPC with different $R$}
    \label{fig:diffR}
\end{figure}

The same examination is conducted for the weight matrix $R$. A scenario is created for hovering the quadrotor at a certain point from the ground with displacement in $x$ and $z$ positions. The evolution of state $z$ and the corresponding inputs $F$, $T_x$, and $T_z$ are illustrated in \autoref{fig:diffR}. The results reveal a relationship that is quite the opposite of that observed with $Q$. Specifically, when more penalties are added to the input, i.e. increasing the value in $R$, the magnitude of the inputs becomes smaller. This can result in worse controller performance due to insufficient control input for the quadrotor. 
Thus, it is crucial to select appropriate weight matrices when designing MPC to ensure optimal performance.

\subsection{Disturbance rejection}
\begin{figure}
    \centering
    \includegraphics[width=\linewidth]{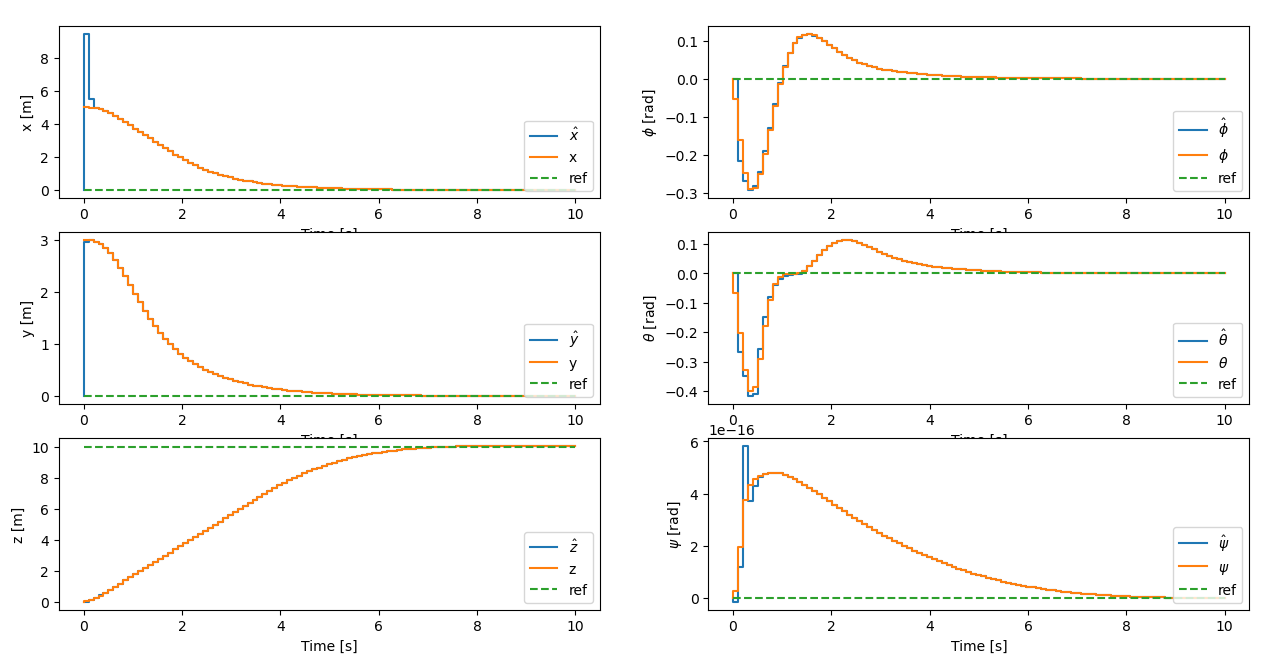}
    \caption{Disturbance rejection, output feedback, reference tracking.}
    \label{fig:dist_of_sim}
\end{figure}

A constant input disturbance is injected into the system acting on the first state $X$, simulating a constant wind shear. An observer is built to estimate the augmented state and feed the current disturbance estimate into the OTS. The initial state of the system $x(0)$, initial observer state $\hat{x}(0)$, and output reference $y_{ref}$ are respectively: 
\begin{align*}
    x(0) &= [5, 3, 0, 0_{9\times 1}]^T\\
    \hat{x}(0) &= [0, 0, 0, 0_{9 \times 1}, \underbrace{0}_{\hat{d}}]^T\\
    y_{ref} &= [0, 0, 10, 0, 0, 0]^T
\end{align*}
Note that here the output is generated through a $C$ matrix that grabs only the first 6 states and maps them without any transformation to the output. \autoref{fig:dist_of_sim} shows the trajectories of the first 6 states, convergence of the observer estimate, and output-free reference tracking.
The time per iteration (i.e. the time for solving the MPC optimization problem once) increases to 0.07s for a horizon $N=10$. This is an absolute increase of 0.02s or relatively $40\%$ compared to the setting where the OTS was solved offline. 

\subsection{Controlling the nonlinear system}
\autoref{sec:intro} shows that the true plant is nonlinear. Hence, our model-based controller and observer suffer from a model mismatch as they were designed using the linearized model. This could be an issue when deploying the designed controller to a real drone. Therefore, the remaining experiment is to show the qualitative difference between driving the linearized model and the original nonlinear model.

\begin{figure}[h!]
    \centering
    \includegraphics[width=\linewidth]{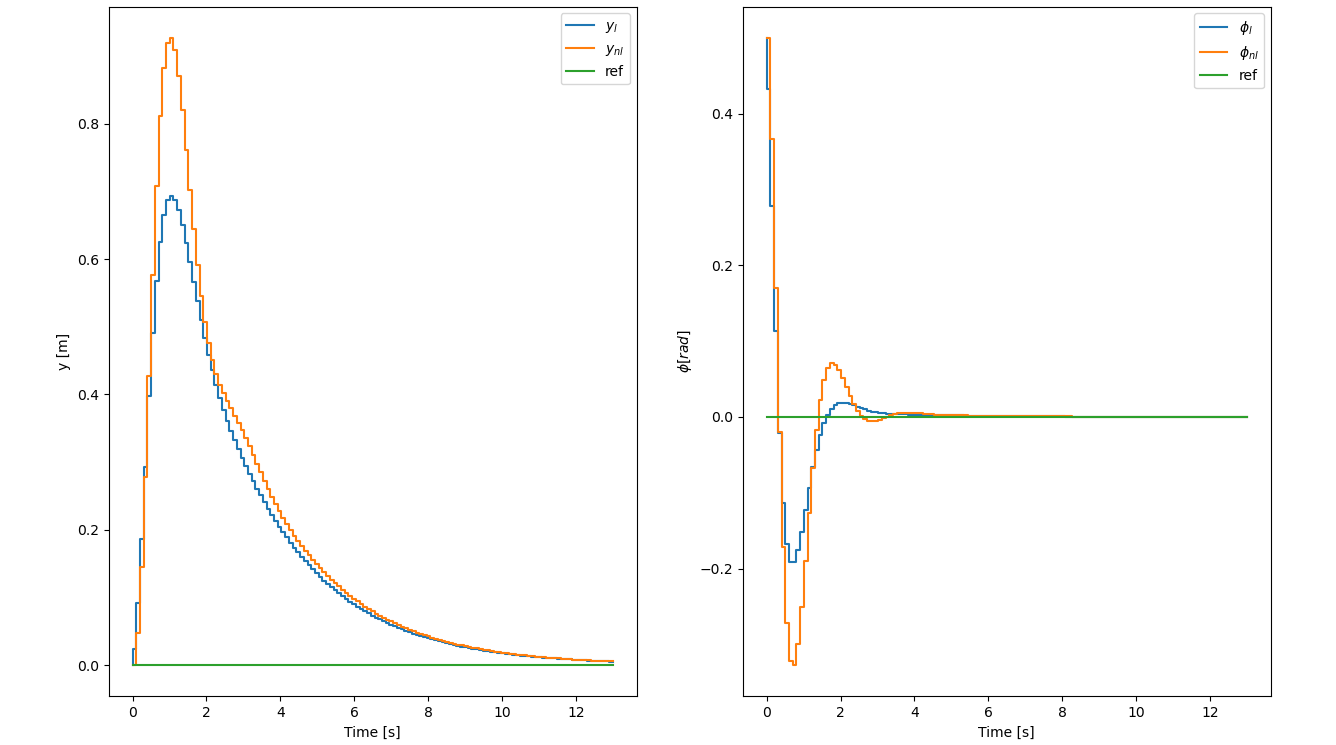}
    \caption{Linear (l) and nonlinear (nl) system behavior comparison with state-feedback}
    \label{fig:nl_sim_state_feedback}
\end{figure}

\begin{figure}[h!]
    \centering
    \includegraphics[width=\linewidth]{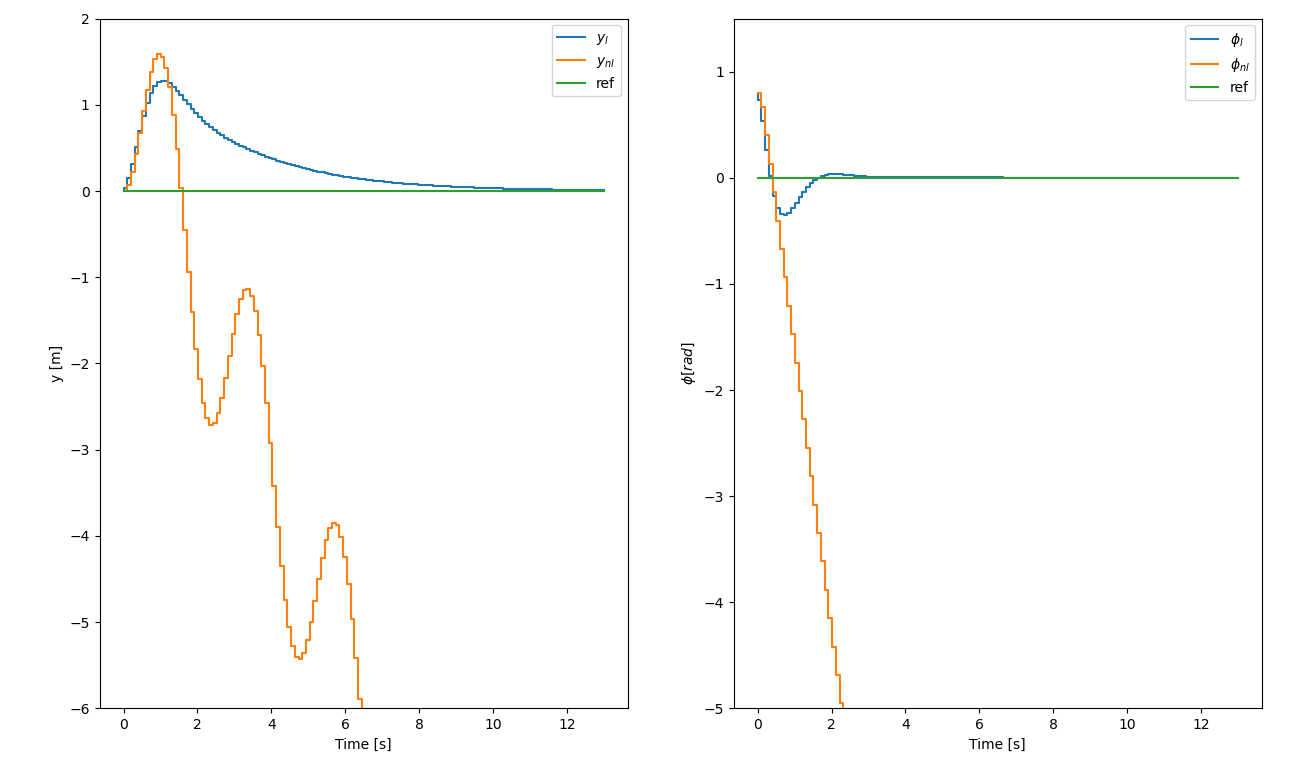}
    \caption{Model mismatch between linear (l) and nonlinear (nl) model, driven with linearized MPC}
    \label{fig:nl_sim_state_feedback_unstable}
\end{figure}

First, we compare the influence of the model mismatch on the performance of the full-state-feedback MPC. \autoref{fig:nl_sim_state_feedback} shows the trajectories of the y-position and roll angle $\phi$. From the figure, it can be observed that the nonlinear system can be stabilized with an initial condition of $\phi(0) = 0.5 rad$. However, compared to the linearized system, the nonlinear system has a larger overshoot, indicating a worse performance.
When increasing the initial value to $\phi(0) = 0.8 rad$, the designed MPC fails to stabilize the nonlinear system as shown in \autoref{fig:nl_sim_state_feedback_unstable}. This comparison indicates that the model mismatch can cause vastly different behavior of the closed-loop system.\\
A similar analysis can be conducted to investigate the influence of error stemming from the model-based observer on the performance of the MPC. Thus, it is crucial to remember these error sources when designing the MPC for a real plant.
% Finally, another error source is the model-based observer which is synthesized using a linearized model of the system. Hence, a similar analysis can be done.

\subsection{Comparison between LQR and MPC}
\begin{figure}[h!]
    \centering
    \includegraphics[width=\columnwidth]{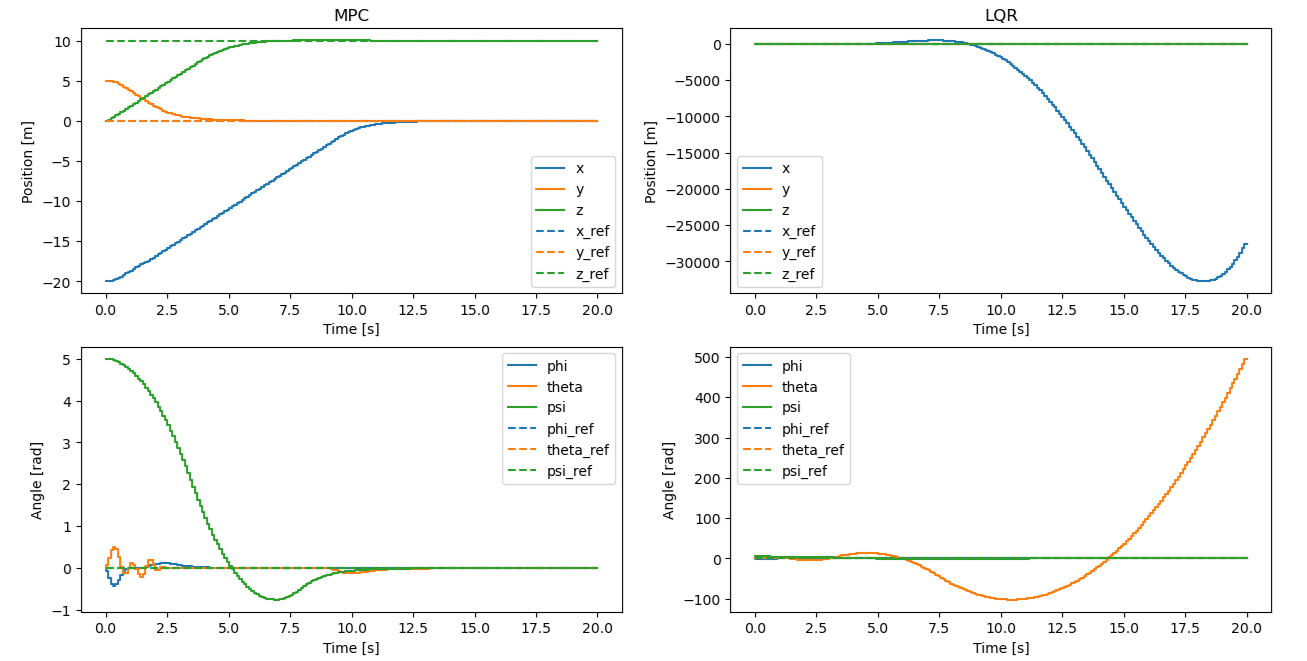}
    \caption{Comparison between the MPC and finite LQR}
    \label{fig:MPCvsLQR}
\end{figure}

The Model Predictive Control (MPC) demonstrates superiority over the Linear Quadratic Regulator (LQR) in two key aspects. Firstly, when considering constraints, the MPC is capable of computing results explicitly, whereas the LQR requires implicit computation. Secondly, the MPC ensures stability for finite horizons $N$, a guarantee that the LQR does not provide. To prove the superiority, the performance of full state-feedback MPC and unconstrained finite-horizon LQR is compared. The process of acquiring gain $\Tilde{K}_{LQR}$ is shown in the Appendix. The weight matrices $Q$ and $R$ are kept identical for comparison. Furthermore, the initial point $x_0$ is chosen out of the terminal set $\mathbb{X}_f$ intentionally. \autoref{fig:MPCvsLQR} shows the simulation result, from which it can be seen that the MPC can stabilize the quadrotor at the hovering point while the LQR cannot. This result further proves the effectiveness of the MPC controller.

%\section{Numerical simulations}

% In this section, we run several numerical simulations where we compare some MPC controllers as well as standard controllers used in Some Application.

% Since in Some Application, the state/output cannot be accurately measured, we assume the presence of a random measurement disturbance...

% Plots... show the effect of shorter/longer control horizon... tuning the cost matrices... Compared with Standard Control, we observe that...

\section{Discussion and Conclusion} \label{sec:Conclusion}
This report examines the application of Model Predictive Control (MPC) in controlling a quadrotor. Two controllers are designed and tested: the full-state-feedback MPC, intended for ideal situations where all states are measurable, and the output-feedback MPC, designed for scenarios where not all states are measurable. Stability analysis is conducted to demonstrate the asymptotic stability of the equilibrium in the closed-loop system. The designed MPCs are validated through numerical simulations in both reference tracking and disturbance rejection (e.g., constant gust of wind).

Despite the advantages of the MPC controller, several concerns must be addressed when applying MPC in real-life scenarios. First, MPC relies heavily on an accurate model, therefore, we analyzed the model error introduced by the linearization. Second, MPC may be computationally heavy; attention must be paid to ensure that each iteration step can be carried out in real time without introducing a time delay. For this, we analyzed how different horizon choices and the OTS influence solving time.

Experiences from deploying MPC in real plants during Robotic's master's RO47007 (mobile manipulator robot) and in the Formula Student Delft (DUT24) racing team highlight this issue. Recent research on learning-based Model Predictive Control and data-driven Model Predictive Control (e.g.~\cite{hewing2020learning}~\cite{torrente2021data}) offers exciting solutions to this problem.

\section{Code}
Please find our code on 
\href{https://github.com/DavidChen0429/MPC-2.0}{this GitHub-repository}.

% Reference
\bibliographystyle{IEEEtran}
\bibliography{mpc_library}

% Appendix
\clearpage
\appendix
\subsection{$\mathbb{X}_f$ Estimation}
\begin{algorithm}[H]
\caption{Estimate Set $\mathbb{X}_f$}
\label{alg:estimate_set_Xf}
\textbf{Result:} $Hx \leq h$ representing $\mathbb{X}_f$
\begin{algorithmic}[1]
\State \textbf{Initialization:}
\State $K :=$ LQR gain
\State $K' := [K; I]$ \Comment{Augmented K}
\State $A_K := \Phi - \Gamma K$ \Comment{CL System Matrix}
\State $s := $ Number of constraints
\State Set $t := 0$
\State \textbf{Iteration:}
\For {$i = 1, 2, \ldots, s$}
\State $x_i^* := \arg\max\limits_{x} \ f_i(K'A_K^{t+1}x)$
\State \hspace{1cm} \text{s.t.} $f_j(K'A_K^k x) \leq 0, \ \forall j \in \{ 1, 2, \ldots, s \}$ 
\State \hspace{4.2cm} $\forall k \in \{ 0, 1, \ldots, t\}$
\EndFor
\If {$f_i(K'A_K^{k+1} x_i^*) \leq 0 \ \forall i \in \{ 1, 2, \ldots, s \}$}
    \State set $t = t^*$
    \State $\mathbb{X}_f := \{ x \in \mathbb{R}^n \mid f_j(K'A_K^k x) \leq 0,$
    \State \hspace{1.5cm} $\forall j \in \{1,2,\ldots,s\}, \ \forall k \in \{0,1,\ldots,t^*\} \}$
    \State Derive and return $H, h$
\Else
    \State set $t := t + 1$ and continue.
\EndIf
\end{algorithmic}
\end{algorithm}

\subsection{Finite LQR Gain Acquisition (Dynamic Programming)} \label{sec:finiteLQR}
To get the gain $L$ of finite LQR, the DPA (Dynamic Programming Algorithm) is used. The math below is done by taking the book~\cite{bertsekas2012dynamic} as a reference. We start our acquisition process with a simple first-dimension system. The DPA problem can be defined as
\begin{itemize}
    \item State Space: $x \in \mathbb{R}$
    \item Input Space: $u \in \mathbb{R}$
    \item Cost Function:
    \begin{itemize}
        \item Stage cost $g(x,u) = qx^2+ru^2$
        \item Terminal cost $G(x) = q_tx^2$
    \end{itemize}
    \item System Dynamic: $X_{t+1} = AX_t+BU_t$
\end{itemize}
with the final goal of minimizing \autoref{eq:finiteLQR}
\begin{equation} \label{eq:finiteLQR}
    J_0^*(x) = \min_{(\mu_t: \mathbb{R} \rightarrow \mathbb{R})_{t=0}^{T-1}} \mathbb{E} \left( \sum_{t=0}^{T-1} qX_t^2 + r \mu_t(X_t)^2 + q_tX_T^2 \Bigg| X_0\right)
\end{equation}
This can be solved by deploying DPA:
\begin{itemize}
    \item \textbf{Initialization} at $t=T$: $J_T(x)=G(x) = q_tx^2, \forall x \in \mathbb{R}$
    \item \textbf{Backward iteration} for $t=T-1,T-2,\ldots,0$
    \begin{equation} \label{eq:DPA2}
        \begin{aligned}
            J_t(x) &= \min_{u \in \mathbb{R}} \left\{ g(x, u) + \mathbb{E}(J_{t+1}(X_{t+1}) \mid X_t, U_t) \right\} \\
            &= \min_{u \in \mathbb{R}} \left\{ qx^2 + ru^2 + \mathbb{E}(J_{t+1}(ax + bu) \mid X_t, U_t) \right\} \\
            &= \min_{u \in \mathbb{R}} \left\{ qx^2 + ru^2 + \mathbb{E}(J_{t+1}(ax + bu)) \right\}, \quad \forall x \in \mathbb{R},
        \end{aligned}
    \end{equation}
    by replacing the terms with the already known definition, \autoref{eq:DPA2} can be rewritten as:
    \begin{equation}
        \begin{aligned}
            J_{T-1}(x) &= \min_{u \in \mathbb{R}} \left\{ qx^2 + ru^2 + \mathbb{E}(q_t(ax + bu)^2) \right\} \\
            &= \min_{u \in \mathbb{R}} \left\{ qx^2 + ru^2 + q_t(ax + bu)^2 \right\},
        \end{aligned}
    \end{equation}
    the minimization problem is a convex quadratic function of $u$. Hence, by using the first-order optimality condition (derivative equals zero), the optimal cost can be acquired as 
    \begin{equation} \label{eq:DPAoptimalCost}
        \begin{aligned}
            0 &= \left. \frac{\partial}{\partial u} \left( qx^2 + ru^2 + q_t(ax + bu)^2 \right) \right|_{u=u^*} \\
            &\Longleftrightarrow 0 = 2(q_tb^2 + r)u^* + 2abq_txu \\
            &\Longleftrightarrow u^* = \mu_{T-1}(x) = -\frac{abq_t}{q_tb^2 + r}x, \quad \forall x \in \mathbb{R}, \\
            &\Longleftrightarrow u^* = l_tx, \quad \forall x \in \mathbb{R}, \\
        \end{aligned}
    \end{equation}
\end{itemize}
By using the induction method, \autoref{eq:DPAoptimalCost} can be proved as a common pattern for all time steps in the prediction horizon $T$. 

This pattern is also shared by the multi-dimension problem where the problem is now defined as:
\begin{itemize}
    \item State Space: $X \in \mathbb{R}^n$
    \item Input Space: $U \in \mathbb{R}^m$
    \item Cost Function:
    \begin{itemize}
        \item Stage cost $g_t(x,u) = X_t^TQX_t+U_t^TRU_t$
        \item Terminal cost $G(x) = X_T^TQ_TX_T$
    \end{itemize}
    \item System Dynamic: $X_{t+1} = A_tX_t+B_tU_t$
\end{itemize}
By using the DPA, the optimal policy is a linear state feedback of the form
\begin{equation}
    \mu_t(x) = L_t x, \forall x \in \mathbb{R}
\end{equation}
where the matrix gains $L_t$ are given by 
\begin{equation}
    L_t = -(B_t^\top K_{t+1} B_t + R_t)^{-1} B_t^\top K_{t+1} A_t, \quad t = T-1, \ldots, 0.
\end{equation}
the gain $K_t$ is positive semi-definite matrices given by
\begin{equation}
    \begin{aligned}
        K_T &= Q_T \\
        K_t &= \mathcal{F}(K_{t+1}), \forall t=T-1,\ldots,0
    \end{aligned}
\end{equation}
where
\begin{equation} \label{eq:finiteLQR_gain}
    \mathcal{F}_t(K) := Q_t + A_t^\top \left(K - K B_t (B_t^\top K B_t + R_t)^{-1} B_t^\top K\right) A_t.
\end{equation}

\autoref{eq:finiteLQR_gain} will converge when $t \rightarrow \infty$ as the unique positive semi-definite solution to the DARE (Discrete Algebraic Riccati Equation), standing out as the main different between the finite LQR and infinite LQR. 

\end{document}